\documentclass[aps,prd,showpacs,nofootinbib,twocolumn,superscriptaddress,preprintnumbers]{revtex4}
\usepackage[english]{babel}
\usepackage{amsmath,amssymb}
\usepackage[dvips]{graphicx}
\usepackage{amsfonts}
\usepackage{stmaryrd}

\usepackage{color}

\renewcommand{\baselinestretch}{1.3}

\newcommand{\CL}   {C.L.}

\newcommand{\eps}  {\varepsilon}
\newcommand{\epp}  {\varepsilon'}
\newcommand{\plumin}[2]{^{+#1}_{-#2}}

\def\2te{2{\theta}}

\def\'#1{\ifx#1i\accent19\i\else\accent19#1\fi}

\def\8{\infty}

\newcommand{\AHEP}{Instituto de F\'{\i}sica Corpuscular --
  C.S.I.C./Universitat de Val{\`e}ncia \\
  Campus de Paterna, Apt 22085,
  E--46071 Val{\`e}ncia, Spain}
\newcommand{\Cinvestav}{Departamento de F\'{\i}sica, Centro de
  Investigaci{\'o}n y de Estudios Avanzados del IPN\\ Apdo. Postal
  14-740 07000 Mexico, DF, Mexico}
\newcommand{\AddrMariam}{%
 II. Institut f\"ur Theoretische Physik, Universit\"at Hamburg,\\
          Luruper Chaussee 149, 22761 Hamburg, Germany}

\begin{document}

\preprint{IFIC/09-30}

\title{Constraining nonstandard neutrino-quark interactions
 with solar, reactor and accelerator data}

\author{F. J. Escrihuela}\email{franesfe@alumni.uv.es}
\affiliation{\AHEP} 

\author{O. G. Miranda}\email{Omar.Miranda@fis.cinvestav.mx}
\affiliation{\Cinvestav}

\author{M.~A.~T\'ortola}\email{mariam.tortola@desy.de}
\affiliation{\AddrMariam}

\author{J. W. F. Valle}\email{valle@ific.uv.es, URL: http://ahep.uv.es/}
\affiliation{\AHEP}

\begin{abstract} 
  We present a reanalysis of nonstandard neutrino-down-quark
  interactions of electron and tau neutrinos using solar, reactor and
  accelerator data.  In addition updating the analysis by including
  new solar data from SNO phase III and Borexino, as well as new
  KamLAND data and solar fluxes, a key role is played in our analysis
  by the combination of these results with the CHARM data. The latter
   allows us to better constrain the axial and axial-vector
  electron and tau-neutrino nonstandard interaction parameters characterizing 
  the deviations from the
  Standard Model predictions.
\end{abstract}


\maketitle

\section{Introduction}
\label{sec:introd}

Current solar neutrino
data~\cite{cleveland:1998nv,Abdurashitov:2009tn,Altmann:2005ix,Kaether:2007zz,fukuda:2002pe,
  Hosaka:2005um,ahmad:2002jz,ahmad:2002ka,
  Ahmed:2003kj,Aharmim:2005gt,Aharmim:2008kc,Arpesella:2008mt,Galbiati:2008zz,Collaboration:2008mr},
in conjunction with reactor data from the KamLAND
experiment~\cite{:2008ee} shows that the neutrino oscillation
mechanism is the correct picture to explain the solar neutrino
physics. Solar neutrino experiments are also sensitive to matter
effects~\cite{wolfenstein:1978ue,mikheev:1985gs}, and the combination
of both solar and KamLAND data determines the so called Large Mixing
Angle (LMA) solution as the correct explanation to the data.  For
example, the LMA solution is quite robust against possible
uncertainties in solar physics, such as noise density fluctuations
originated by radiative zone magnetic
fields~\cite{Schaefer:1987fr,Krastev:1991hc,Loreti:1994ry,nunokawa:1996qu,Burgess:2003fj,burgess:2002we,Burgess:2003su,Fogli:2007tx}. Likewise,
the LMA solution is also stable with respect to the possible existence
of sizeable convective zone magnetic
fields~\cite{miranda:2000bi,Miranda:2003yh}, that could induce
spin-flavor neutrino
conversions~\cite{schechter:1981hw,akhmedov:1988uk}.
In all these cases, the KamLAND data play a crucial role in
establishing that nonstandard effects can only play a subleading
role~\cite{pakvasa:2003zv}, their amplitude being effectively
constrained.

However, while constrained by the solar and KamLAND data in an
important way, neutrino nonstandard interactions (NSI) still provide
an important exception to the robustness of the neutrino oscillation
interpretation~\cite{Friedland:2004pp,Guzzo:2004ue}.
Indeed, it has been found that they might even shift the solution to
the so--called dark side region of the neutrino parameter
space~\cite{Miranda:2004nb}. 

Given the envisaged precision expected in upcoming oscillation
studies~\cite{Bandyopadhyay:2007kx}, one needs to further scrutinize
the possible role of NSI~\cite{Nunokawa:2007qh}.
The intrinsic importance of NSI stems from the fact that they are
characteristic features of theories of neutrino
mass~\cite{Valle:2006vb} and that their magnitude provides important
guidance in order to distinguish the simplest high-scale seesaw
models~\cite{gell-mann:1980vs,yanagida:1979,mohapatra:1980ia,schechter:1980gr,Schechter:1981cv}
from those seesaw scenarios based at low-scale physics, such as the
inverse~\cite{mohapatra:1986bd,Bazzocchi:2009kc} or linear seesaw
mechanisms~\cite{Malinsky:2005bi}, as well as radiative models of
neutrino mass~\cite{babu:1988ki,zee:1980ai,AristizabalSierra:2007nf}.

In this paper we reanalyze the robustness of the oscillation
interpretation of the solar neutrino data in the presence of
nonstandard interactions. Besides all the solar neutrino data used in
our previous study~\cite{Miranda:2004nb}, here we take into account
new solar data from SNO phase III~\cite{Aharmim:2008kc}, the first
real-time measurements of the solar Beryllium flux at
Borexino~\cite{Arpesella:2008mt}, as well as the new and more precise
KamLAND data~\cite{:2008ee}.  We have also considered in our
calculations the new solar fluxes and uncertainties from the updated
Standard Solar Model (SSM)~\cite{PenaGaray:2008qe}.
We show explicitly that the degenerate solution in the dark
side region still remains plausible even after inclusion of these
new data.
Besides updating the analysis of nonstandard neutrino-down-quark
interactions, we stress the key role is played by the combination of
these results with the measurements of the electron-neutrino-quark
cross section at the CHARM accelerator experiment. 
Although it is sensitive only to the interactions of electron
neutrinos, when combined with solar and KamLAND data, the latter
allows us to improve the determination of the tau-neutrino nonstandard axial
and vector couplings

In what follows we will focus on nonstandard interactions that can
be parametrized with the effective low--energy neutral currents
four--fermion operator \footnote{A recent study of CC non-standard
  interactions has been given in Ref.~\cite{Biggio:2009nt}.} :
\begin{equation}
    \mathcal{L}_{NSI} = 
    - \varepsilon_{\alpha\beta}^{f P} 2\sqrt{2}G_F \left(
    \bar{\nu}_\alpha\gamma_\mu L \nu_\beta
\right)
\left(\bar{f} \gamma^\mu P f \right),
\label{lagrang}
\end{equation}
where P = L, R and $f$ is a first generation fermion: $e,u,d$. The
coefficients $\eps_{\alpha\beta}^{fP}$ denote the strength of the NSI
between the neutrinos of flavours $\alpha$ and $\beta$ and the
P--handed component of the fermion $f$.
For definiteness, we take for $f$ the down-type quark.  However, one
can also consider the presence of NSI with electrons and up and down
quarks simultaneously. Current and expected limits for the case of NSI
with electrons have been reported in the
literature~\cite{deGouvea:2006cb,Barranco:2005ps,Bolanos:2008km}.
Here we confine ourselves to NSI couplings involving only
electron and tau neutrinos.
This approximation is in principle justified in view of the somewhat
stronger constraints on $\nu_\mu$ interactions, for a discussion see
Ref.~\cite{Davidson:2003ha,Biggio:2009kv,Berezhiani:2001rs}.

Nonstandard interactions may in principle affect neutrino
propagation properties in matter as well as detection cross sections
and in certain cases they can also modify the assumed initial
fluxes\footnote{We assume a class of models of neutrino mass where NSI
leave the solar and reactor neutrino fluxes unaffected.}.
NSI effects in neutrino propagation affect the analysis of data from
solar neutrino experiments and to some extent also KamLAND, through
the vectorial NSI couplings $\eps_{\alpha\beta}^{d V} =
\eps_{\alpha\beta}^{d L} + \eps_{\alpha\beta}^{d R}$. On the other
hand detection shows sensitivity also to the axial NSI couplings
$\eps_{\alpha\beta}^{d A} = \eps_{\alpha\beta}^{d L} -
\eps_{\alpha\beta}^{d R}$ in the SNO experiment. These points will be
analyzed in detail in Section~\ref{sec:solar-kamland-data}, after a
brief discussion of the experimental data included in our study. In
Section~\ref{sec:constraints-nsi-from} we will focus on the study of
the non-universal nonstandard interactions, combining the results of
the CHARM experiment together with our solar analysis in order to
obtain a new constraint for the tau neutrino nonstandard-interaction
with d-type quark. Finally we will conclude in
Section~\ref{sec:conclusions}.  

\section{Sensitivity of solar and KamLAND data to NSI}
\label{sec:solar-kamland-data}

Here we will adopt the simplest approximate two--neutrino picture,
which is justified in view of the stringent limit on
$\theta_{13}$~\cite{Schwetz:2008er} that follows mainly from reactor
neutrino experiments~\cite{apollonio:1999ae}.

\subsection{Data}
\label{sec:data}

In this subsection we will describe the input data required to analyze
the sensitivity of solar and KamLAND neutrino data to the presence of
NSI. This will include not only the experimental data samples by all
the detectors considered, but also the theoretical predictions
required to simulate the solar neutrino production prescribed by the
SSM.

Concerning the solar neutrino data, we have included in our analysis
the most recent results from the radiochemical experiments
Homestake~\cite{cleveland:1998nv}, SAGE~\cite{Abdurashitov:2009tn} and
GALLEX/GNO~\cite{Altmann:2005ix,Kaether:2007zz} , the zenith-spectra
data set from Super-Kamiokande I~\cite{fukuda:2002pe, Hosaka:2005um},
as well as the results from the two first phases of the SNO
experiment~\cite{ahmad:2002jz,ahmad:2002ka,
  Ahmed:2003kj,Aharmim:2005gt}.
The main updates with respect to our previous
work~\cite{Miranda:2004nb} is the inclusion of the data from the third
phase of the SNO experiment~\cite{Aharmim:2008kc}, where $^3$He
proportional counters have been used to measure the neutral current (NC)
component of the solar neutrino flux, and the latest measurement of
the $^7$Be solar neutrino rate performed by the Borexino
collaboration~\cite{Arpesella:2008mt,Galbiati:2008zz}.

In our analysis we use the solar neutrino fluxes and uncertainties
given by the latest version of the SSM~\cite{PenaGaray:2008qe}. The
latter provides an improved determination of the neutrino flux
uncertainties, mainly thanks to the improved accuracy on the
$^3$He-$^4$He cross section measurement and to the reduced systematic
uncertainties in the determination of the surface composition of the
Sun. In Ref.~\cite{PenaGaray:2008qe} two different solar model
calculations are presented, corresponding to two different
measurements of the solar metal abundances. For our analysis we have
chosen the model corresponding to a higher solar metallicity,
BPS08(GS), although we have checked that the use of the lower metallicity
model BPS08(AGS) does not change our results substantially.

The KamLAND experiment observes the disappearance of reactor
antineutrinos over an average distance of 180 km. Given that, in their
way to the detector, reactor neutrinos can only traverse the most
superficial layers of the Earth, the resulting Earth matter effects
are almost negligible. The same applies also to the nonstandard
interactions we are considering. However, for consistency with our
analysis of solar neutrino data, in our simulation of the KamLAND
experiment, we have included the effect of NSI over the antineutrino
propagation. In particular, we have considered that neutrinos travel
through a layer of constant matter density equal to the terrestrial
crust density ($\sim$ 2.6 g$\cdot$cm$^{-3}$).
Here we have used the latest data release from the KamLAND reactor
experiment~\cite{:2008ee}, with a total exposure of 2881~ton$\cdot$yr,
which brings in a big statistical improvement with respect to the
previous data reported by the Collaboration~\cite{Araki:2004mb}. We
have restricted our analysis to the energy range above 2.6~MeV where
the contributions from geo-neutrinos is less important.  As we will
see in the next section, the inclusion of the new KamLAND data will be
very important for the improvement of the results, given the good
precision achieved in the determination of the oscillation neutrino
parameters.

\subsection{Effects in neutrino propagation}

We first reanalyse the determination of the oscillation parameters in the presence of nonstandard interactions.
The Hamiltonian describing solar neutrino evolution in the presence of NSI contains, in addition to the standard oscillations term, 
\begin{equation}
\left(\begin{array}{lc}
-\frac{\Delta m^2}{4E}\cos 2\theta + \sqrt{2}\, G_F N_e ~~ & 
\frac{\Delta m^2}{4E}\sin 2\theta \\
~~~\frac{\Delta m^2}{4E}\sin 2\theta &
\frac{\Delta m^2}{4E}\cos 2\theta
\end{array} \right) \, ,
\label{eq:msw-hamiltonian}
\end{equation}
a term $H_\mathrm{NSI}$, accounting for an effective potential induced by the NSI with matter, which may be written as:
\begin{equation}
    H_\mathrm{NSI} = \sqrt{2} G_F N_d
    \left( \begin{array}{cc}
        0 & \varepsilon \\ \varepsilon & \varepsilon'

    \end{array}\right) \,.
    \label{eq:nsi-hamiltonian}
\end{equation}

Here $\varepsilon$ and $\varepsilon'$ are two effective parameters
that, according to the current bounds discussed above
($\eps_{\alpha\mu}^{f P} \sim 0$), are related with the vectorial
couplings which affect the neutrino propagation by\footnote{
    For the derivation of the effective couplings in the general
    three-neutrino framework see Ref.\cite{Guzzo:2001mi}.}:
\begin{equation}
    \eps = - \sin\theta_{23}\,\eps_{e\tau}^{d V} \qquad
    \epp = \sin^2\theta_{23}\,\eps_{\tau\tau}^{d V} -
    \eps_{ee}^{d V} 
    \label{eff-coup}
\end{equation}
The quantity $N_d$ in Eq.~(\ref{eq:nsi-hamiltonian}) is the number
density of the down-type quark along the neutrino path, and $\theta_{23}$ is the atmospheric neutrino mixing angle.

From Eqs.~(\ref{eq:msw-hamiltonian}) and (\ref{eq:nsi-hamiltonian}) one
sees that the solar neutrino mixing angle in the presence of
nonstandard interactions is given by the following expression:
\begin{equation}
    \cos 2\theta_m=
    \frac{\Delta m^2\cos 2\theta - 2\sqrt{2}\,E G_F (N_e 
      - \varepsilon' N_d)}
    {\left[\Delta m^2\right]_{matter}} ,
    \label{cos-mat-nsi}
\end{equation}
where
\begin{multline}
    \left[
        \Delta m^2
    \right]_{matter}^2 = 
    \left[
        \Delta m^2\cos 2\theta -
        2\sqrt{2}\,E G_F (N_e - \varepsilon' N_d) 
    \right]^2 
    \\
    +
    \left[\Delta m^2
        \sin 2\theta + 4\sqrt{2}\,\varepsilon\, E G_F N_d 
    \right]^2
\end{multline}
Therefore, and as discussed in Ref.\cite{Miranda:2004nb}, there exists a
degeneracy between the non-universal coupling $\varepsilon'$ and the
neutrino mixing angle $\theta$ which makes possible to explain the
solar neutrino data for values of the vacuum mixing angle in the dark
side ($\theta >\pi/4$), for large enough values of $\varepsilon'$:
\begin{equation}
\eps' > \frac{2\sqrt{2}\,E G_F N_e + \Delta m^2 |\cos 2\theta|} {2\sqrt{2}\, E G_F N_d}.
\end{equation}
For instance, for the typical values of the solar neutrino
  energies and matter densities one has $\eps' \gtrsim 0.6$.  Indeed,
as we showed in~\cite{Miranda:2004nb}, the effect of NSI on solar
neutrino propagation implies the presence of an additional LMA-D
solution whose status we now reanalyse in the light of new data.

%% vvv

We now turn to the combined solar + KamLAND analysis.  Following the
considerations above, we have performed a new analysis of all the
solar neutrino data discussed in Section \ref{sec:data} combined with
the recent KamLAND result~\cite{:2008ee}. 
\begin{figure*}[t]
\begin{center}
\includegraphics[width=.8\textwidth,angle=0]{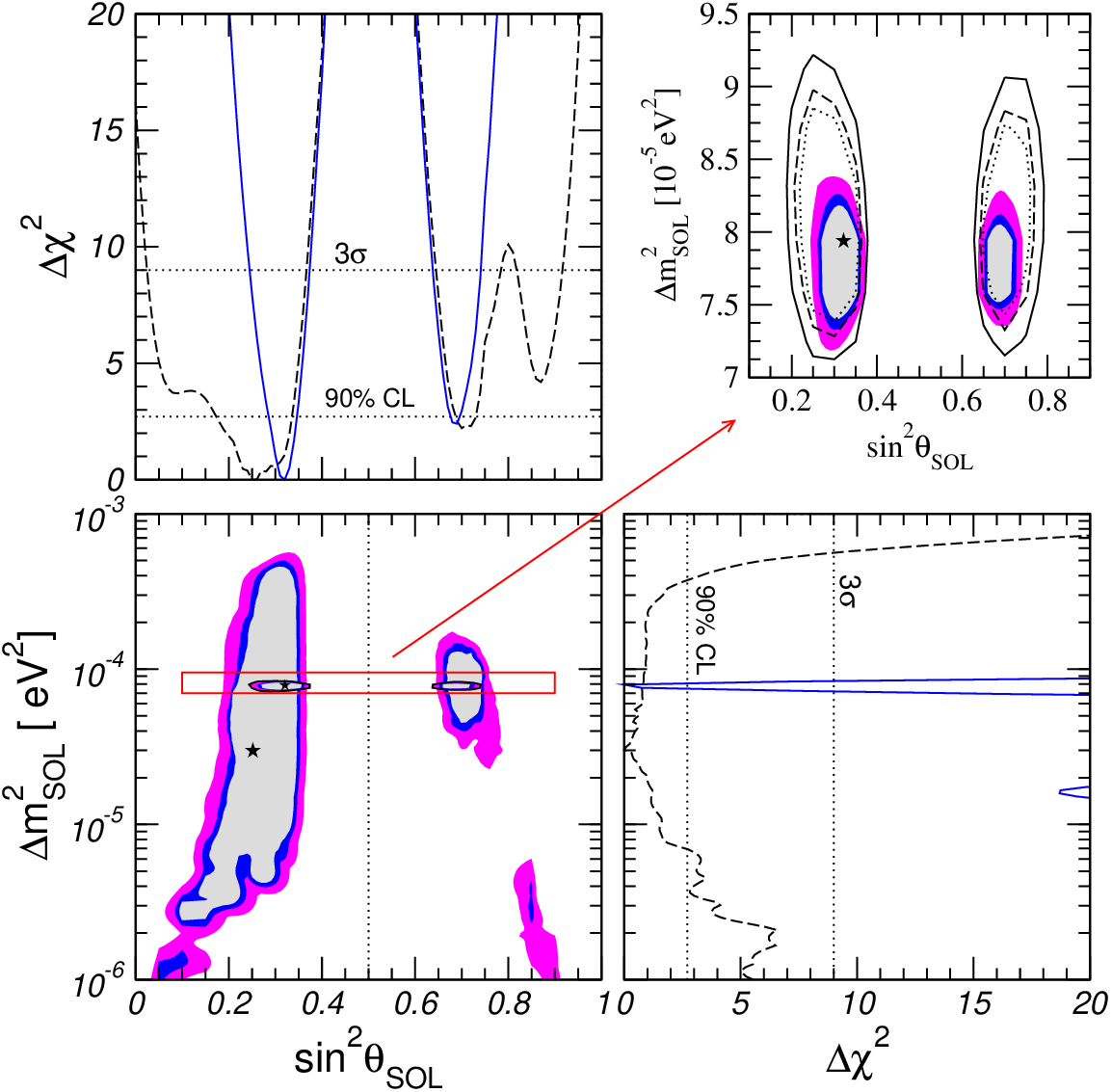}
\caption{The left-bottom panel indicates 90\%, 95\% and 99\% \CL\
  allowed region from the solar and solar + KamLAND combined analysis,
  the result of which is presented as a zoom in the top-right panel,
  where the shaded regions show the result of the current analysis
  while the lines indicate the earlier regions~\cite{Miranda:2004nb}
  for comparison.  One sees that the dark side solution is still
  allowed in the presence of nonstandard interactions.  The other two
  panels indicate the corresponding $\chi^2$ projections. In all
    cases, we marginalize over the NSI parameters $\eps$ and $\epp$. }
\label{fig:updated}
\end{center}
\end{figure*}
The main result is shown in Fig.~\ref{fig:updated}.  There, we plot
the allowed regions (90, 95 and 99 \% \CL) in the solar neutrino
oscillation parameter space ($\sin^2\theta_{\mathrm{SOL}}, \Delta
m^2_{\mathrm{SOL}}$) obtained in the analysis of solar and solar +
KamLAND neutrino data, after marginalizing over the NSI
  parameters in our 4-dimensional $\chi^2$ analysis: 
  $\chi^2$($\sin^2\theta_{\mathrm{SOL}}$, $\Delta m^2_{\mathrm{SOL}}$,
  $\varepsilon$, $\varepsilon'$). The $\Delta \chi ^2$ profiles as a
function of each parameter are also shown.
One can see that the region in the so-called dark side of the neutrino
parameters~\cite{Miranda:2004nb} remains even after the inclusion of
the new data.  Note however that its status is somewhat worse than
previously.
  In contrast, as seen in the figure, the other solutions LMA-0 and
  LMA-II~\cite{Miranda:2004nb} which were present before have
  disappeared as a result of the new KamLAND data, that now provide a
  very precise measurement of $\Delta m^2$.

\begin{figure*}
\begin{center}
  \includegraphics[height=6cm,width=.6\textwidth,angle=0]{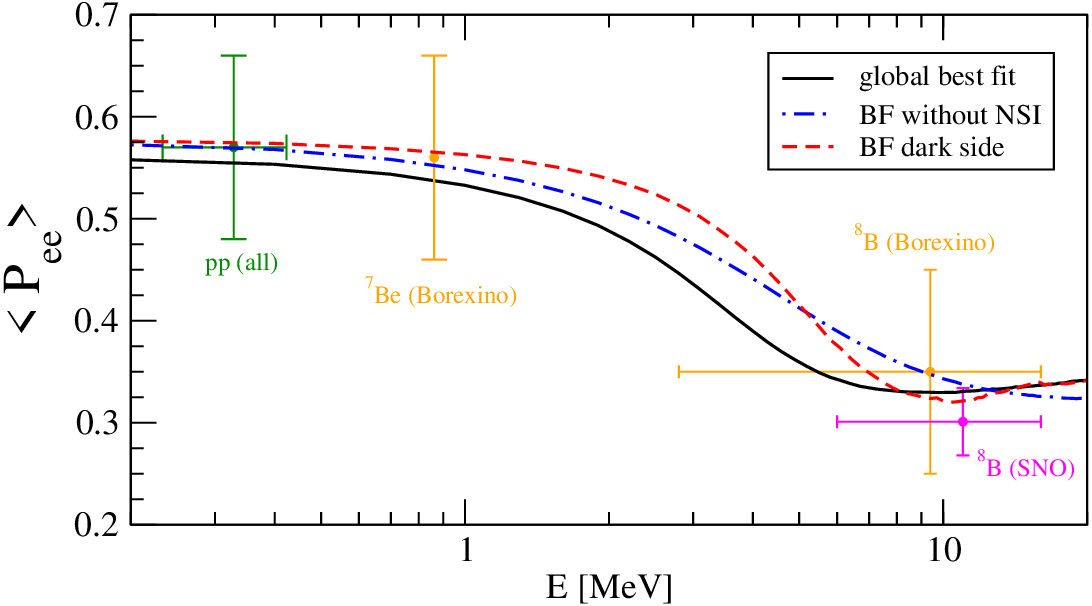}
  \caption{Neutrino survival probabilities averaged over the
    $^8$B neutrino production region for three reference points (see
    text for details). Lines are compared with the experimental rates
    for the pp neutrino flux (from the combination of all solar experiments),
    the 0.862 MeV $^7$Be line (from Borexino) and two estimated values
    of the $^8$B neutrino flux from Borexino and SNO (third
    phase). Here, vertical error bars correspond to the experimental
    errors, while the horizontal ones indicate the energy range
    observed at the experiment. }
\label{fig:pee}
\end{center}
\end{figure*}

  In order to better understand the results obtained, we plot at 
  Fig.~\ref{fig:pee} the neutrino survival probabilities for different reference
  points. First we have considered the global best fit point from the
  combined solar + KamLAND analysis, labeled as "global best fit" in
  the figure, with the following parameter values:
  ($\sin^2\theta_{\mathrm{SOL}}$, $\Delta m^2_{\mathrm{SOL}}$,
  $\varepsilon$, $\varepsilon'$) = (0.32, 7.9$\times$ 10$^{-5}$
  eV$^2$, -0.15, -0.10).
  We have also considered the best fit point in the absence of NSI:
  (0.30, 7.9$\times$ 10$^{-5}$ eV$^2$, 0.00 ,0.00), labeled as "BF
  without NSI", and allowed with a $\Delta \chi^2$ = 2.7, and finally
  the best fit point in the ``dark side'' of the oscillation
  parameters (labeled as "BF dark side") with (0.70, 7.9$\times$
  10$^{-5}$ eV$^2$, -0.15, 0.95) and $\Delta \chi^2$= 2.9.
  As we see all points are in perfect agreement with the low energy
  (pp and $^7$Be) measurements. Concerning the most energetic boron
  neutrinos, where matter effects are more important, and as a
    result there is a higher NSI--sensitivity of the corresponding
    profiles, the presence of NSI provides an slightly better
  agreement with the data than the standard one without NSI, mainly
  thanks to the flatter spectrum predicted above 5 MeV. The
  ``dark-side'' solution also gives predictions for the survival
  probability of boron neutrinos which are compatible with the
  experimental results.
  From the different predictions obtained for these 3 reference points
  above a few MeV, one sees that this region could be crucial in order
  to break the degeneracy among the various solutions. Therefore, a
  better measurement of the boron neutrino flux with a lower threshold
  (like the ones expected from Super-K-III and
  SNO~\cite{Takeuchi:2008zz,SNO-LETA}) will be of great help.
  On the other hand, a very precise measurement of the pep and
  beryllium neutrino fluxes may also contribute to lift the degeneracy
  between the standard and dark-side solutions.

\begin{figure*}
\begin{center}
\includegraphics[height=6cm,width=0.7\textwidth,angle=0]{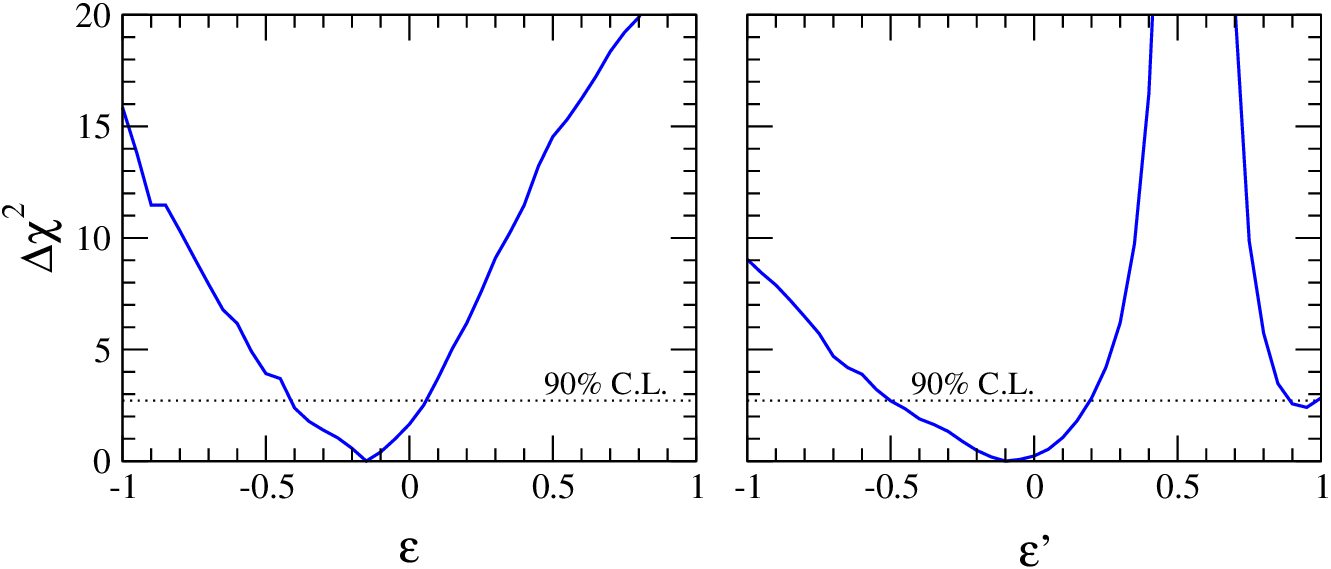}
\caption{Bounds on $\varepsilon$ and $\varepsilon'$ from the combined
  analysis of Solar plus KamLAND neutrino data. One sees that the
  $\varepsilon$ parameter is now constrained to only one region while
  for the case of $\varepsilon'$ there are two possible regions, one
  corresponding to the standard light side solution and the other one
  (less favored) to the so called dark side region.  }
\label{fig:eps-epsp}
\end{center}
\end{figure*}

  By analysing the goodness of the neutrino oscillation solutions in
  the presence of NSI one can also constrain the NSI parameters
  $\varepsilon$ and $\varepsilon '$.  
  In order to do this we first marginalize our 4-parameter
    $\chi^2$ analysis with respect to the remaining three neutrino
    parameters.
  The results are shown in Fig.~\ref{fig:eps-epsp}.   One can see
  that the new data allow us to constraint $\varepsilon$, the flavor
  changing parameter, while for the flavor conserving case there is
  still room for relatively large values of $\varepsilon'$ that
  correspond to the solution in the dark side of the neutrino
  oscillation parameters. These are the bounds we obtained at the 90\% \CL\: 
\begin{eqnarray}
\label{eq:eps-lim0}
 -0.41 & < \varepsilon  < &  0.06  \\ 
 -0.50 <  \varepsilon'  < 0.19  & \&  & 0.89< \varepsilon' < 0.99 
\end{eqnarray}

The above limits are in good agreement with the forecast made in Fig. 3 of
\cite{Miranda:2004nb}, assuming the best possible determination of the
neutrino mixing parameters due to KamLAND . In fact, they are even a
bit better than expected from the improvement of KamLAND data
only. The reason for this is the subsequent improvement of solar data,
which slightly improved their sensitivity to the nonstandard
interactions.

For the flavour-changing effective coupling $\eps$, one can use the
first expression in Eq.~(\ref{eff-coup}) to translate the bound
obtained in Eq.~(\ref{eq:eps-lim0}) into a limit over the vectorial
coupling $\eps_{e\tau}^{dV}$:
\begin{equation}
-0.08 < \eps_{e\tau}^{dV} < 0.58~~~~~~~(90\%~\rm{\CL})
\end{equation}
where we have used the best fit value for the atmospheric mixing
angle~\cite{Schwetz:2008er}. So far, the strongest limit on this
parameter is $|\eps_{e\tau}^{dV}| < 0.5$ \cite{Davidson:2003ha}. From
our analysis, we see that solar neutrino data are not only sensitive
to the sign of this coupling but also we have improved the lower
bound.

 \subsection{Effects in neutrino detection}

 The presence of nonstandard interactions can also affect the
 detection processes at some experiments. In particular, the cross
 section for the neutral current detection reaction at SNO:
\begin{equation}
\nu + d \to \nu + p + n
\end{equation}
is proportional to $g_A^2$, where $g_A$ is the coupling of the
  neutrino current to the axial isovector hadronic 
current~\cite{Bahcall:1988em}. Therefore, the presence of an axial
nonstandard coupling would give rise to an extra contribution to the
NC signal at the SNO experiment. This nonstandard contribution can be
parametrised in the following way~\cite{Davidson:2003ha}:
\begin{equation}
\phi_{NC} \sim f_B (1 + 2\eps_A)~,
\label{eq:nc_epsa}
\end{equation}
where terms of order $\eps_A^2$ have been neglected. Here $f_B$
  denotes the boron neutrino flux, and the effective axial coupling
  $\eps_A$ is defined as in~\cite{Davidson:2003ha}:
\begin{equation}
\eps_A = -\sum_{\alpha = e,\mu,\tau}\left< P_{e\alpha} \right> _{NC} \eps_{\alpha\alpha}^{d A}~,
\label{eq:epsa}
\end{equation}
once the nonstandard axial couplings with up-type quarks are set to zero. Note that $\eps_{\alpha\alpha}^{d A} = \eps_{\alpha\alpha}^{d L} -
\eps_{\alpha\alpha}^{d R}$, denoting the couplings entering in the
effective Lagrangian shown in Eq.~(\ref{lagrang}). Thus, $\eps_A$ is
independent of the effective couplings $\eps$ and $\epp$ defined in
Eq.~(\ref{eff-coup}).  
So far we have assumed in our analysis that $\eps_A = 0$. This
assumption is well justified due to the good agreement between the SNO
NC measurement: $\phi_{NC}^{SNO}$ = $5.54
\plumin{0.33}{0.31}$~(stat) $\plumin{0.36}{0.34}$~(syst) $\times$
$10^6$ $cm^{-2} s^{-1}$ \cite{Aharmim:2008kc} and the SSM prediction
for the boron flux $f_B$ = $5.94 \pm 0.65$ $\times$ $10^6$
$cm^{-2} s^{-1}$ \cite{PenaGaray:2008qe}.  However, we now relax
this assumption by including the effect of the new parameter $\eps_A$.

\begin{figure*}
\begin{center}
\includegraphics[width=0.45\textwidth,angle=0]{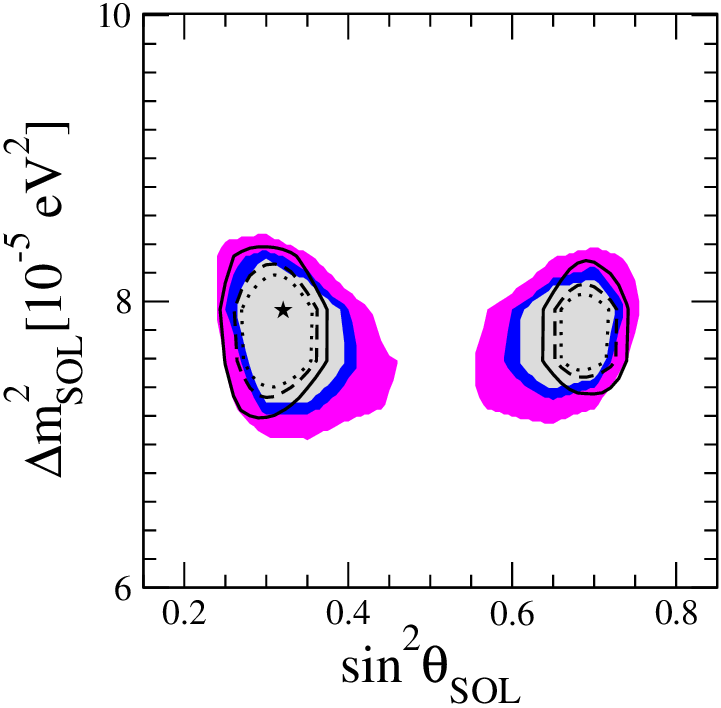}
\includegraphics[width=0.46\textwidth,angle=0]{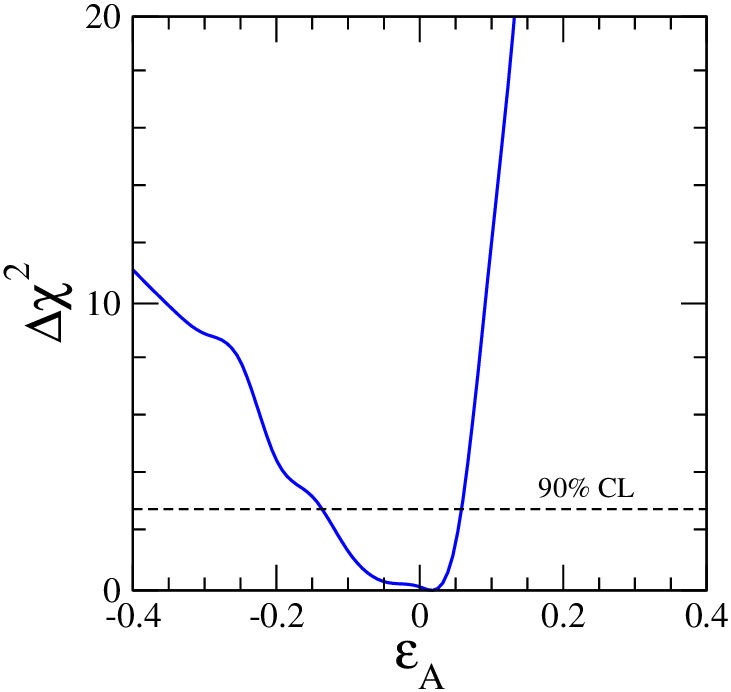}
\caption{The shaded regions in the left panel show the updated
  analysis of all solar neutrino data combined with recent KamLAND
  results in the presence of the axial NSI coupling. The lines show,
  for comparison, the updated regions obtained only with the vector
  NSI couplings.  We show in the right panel the $\chi^2$ profile with
  respect to the axial NSI parameter obtained using all solar neutrino
  data combined with KamLAND .  One sees how the data prefer $\eps_A
  \sim 0$.  }
\label{fig:solar}
\end{center}
\end{figure*}

The results obtained in a generalized 5-parameter-analysis which takes
into account the presence of a non-zero axial component of the NSI are
summarized in Fig.  \ref{fig:solar}.  
In the left panel we compare the allowed regions at 90, 95 and
99\% \CL\ obtained in the full 5-parameter-analysis (filled/colored
regions) with the ones obtained in the previous section, neglecting
the effect of the axial NSI couplings (hollow lines). One
sees that both analysis are consistent, though, as expected, the
inclusion of the axial parameter in the analysis somewhat extends the allowed
region. In the right panel we show the $\Delta \chi^2$ profile as
a function of the effective axial coupling $\eps_A$. There, one
sees that the neutrino data clearly prefers $\eps_A \sim 0$,
thanks to the good agreement between the predicted boron neutrino
flux and the NC observations at SNO, as stated above.  We obtain
we following allowed range at 90\% \CL\:
\begin{equation}
-0.14 < \varepsilon_A < 0.06 
\label{eq:epsa_bound}
\end{equation}

Using Eq.~(\ref{eq:epsa}), the above bound on the effective axial
coupling $\varepsilon_A$ can be translated into individual bounds on
the NSI parameters $\eps_{\alpha\alpha}^{d A}$.  Since we are
neglecting the nonstandard interactions of the muon neutrino, this
formula depends only on the probabilities $\left<P_{ee} \right>
_{NC}$, $\left<P_{e\tau} \right> _{NC}$ and on the NSI couplings
$\eps_{ee}^{d A}$ and $\eps_{e\tau}^{d A}$. From the recent values
for the average probabilities reported by SNO~\cite{Aharmim:2008kc}:
\begin{eqnarray}
\left<P_{ee} \right> _{NC} = 0.30 \pm 0.03 \nonumber \\
\left<P_{e\tau} \right> _{NC} = 0.35 \pm 0.02 \,
\label{eq:ncprobs}
\end{eqnarray}
one gets an allowed region in the parameter space ($\eps_{ee}^{d A}$,
$\eps_{e\tau}^{d A}$), represented (at the 90\% \CL) as a diagonal
band in Fig.~\ref{fig:combined} in the next section. There, it will
used combined with neutrino laboratory data to obtain improved
constraints on the neutrino NSI couplings.

\section{Constraints on non-universal NSI}
\label{sec:constraints-nsi-from}

In addition of solar+ KamLAND, laboratory experiments measuring
neutrino-nucleon scattering show sensitivity to neutrino
non-standard interactions on d-type-quarks. In particular, here we
will combine the results of the accelerator experiment CHARM
together with the ones in Sec.~\ref{sec:solar-kamland-data} in order
to obtain stronger constraints on the NSI parameters. Given the
sensitivity of the considered experiments to different NSI
parameters, we have been forced to simplify the analyses reducing
the number of parameters by focusing on the flavour-conserving
non-universal nonstandard couplings. Within this approximation, the
parameters relevant for each experiment are shown in Table
\ref{table:experiments}.  

\begin{table}[h!]
 \begin{tabular}{|l|c|c|c|c|}
\hline
Data  & $\eps_{ee}^{d V}$ & $\eps_{\tau\tau}^{d V}$ &
                  $\eps_{ee}^{d A}$ & $\eps_{\tau\tau}^{d A}$ \\ \hline
Solar propagation & \checkmark & \checkmark &  &  \\
Solar NC detection  &  &  & \checkmark & \checkmark \\
KamLAND  propagation & \checkmark & \checkmark &  &  \\
CHARM   detection & \checkmark & & \checkmark &  \\
\hline
 \end{tabular}
 \caption{Sensitivity of neutrino experiments to flavor conserving NSI
parameters}
\label{table:experiments}
\end{table}

\subsection{CHARM}

We now turn to the analysis of CHARM data. CHARM was an accelerator
experiment measuring the ratio of the neutral current to the charge
current cross section for electron (anti)neutrinos off quarks. We have
used the results reported by the CHARM experiment for the $\nu_e q \to
\nu q$ cross section. In particular the experiment measured the
relation \cite{Dorenbosch:1986tb},
\begin{eqnarray}
\nonumber
R^e & =  & \frac{\sigma(\nu_e N \to \nu_e X) + \sigma(\bar\nu_e N \to \bar\nu X)}{\sigma(\nu_e N \to e X) +\sigma(\bar\nu_e N \to \bar{e} X) } \\
& = & (\tilde{g}_{Le})^2 +  (\tilde{g}_{Re})^2 = 0.406 \pm 0.140
\label{Re}
\end{eqnarray}

The most general expression for $(\tilde{g}_{L,Re})^2$ including all types of 
NSI parameters is given by 

\begin{widetext}
\begin{eqnarray}
\label{eq:g_Le}
(\tilde{g}_{Le})^2 = (g_L^u + \varepsilon_{ee}^{uL})^2 + \sum_{\alpha \neq e} |\varepsilon_{\alpha e}^{uL}|^2 + (g_L^d + \varepsilon_{ee}^{dL})^2 + \sum_{\alpha \neq e} | \varepsilon_{\alpha e}^{dL}| \\
\label{eq:g_Re}
(\tilde{g}_{Re})^2 = (g_R^u + \varepsilon_{ee}^{uR})^2 + \sum_{\alpha \neq e} |\varepsilon_{\alpha e}^{uR}|^2 + (g_R^d + \varepsilon_{ee}^{dR})^2 + \sum_{\alpha \neq e} | \varepsilon_{\alpha e}^{dR}| 
\end{eqnarray}

\end{widetext}
Here we are interested in the flavor conserving $d$-type quark
interaction. Then, we will neglect all flavor-changing nonstandard
contributions, implying that $\varepsilon_{\alpha e}^{qL} = 0$,
as well as nonstandard couplings to the $u$-type quark, so that
$\varepsilon_{ee}^{uP} = 0$. In this case our simplified expression
for Eqs.~(\ref{eq:g_Le}) and (\ref{eq:g_Re}) would be
\begin{eqnarray}
(\tilde{g}_{Le})^2 = (g_L^u)^2 + (g_L^d +
  \frac{1}{2}(\varepsilon_{ee}^{dV} + \varepsilon_{ee}^{dA}))^2
  \\
(\tilde{g}_{Re})^2 = (g_R^u)^2 + (g_R^d + \frac{1}{2}
  (\varepsilon_{ee}^{dV} - \varepsilon_{ee}^{dA}))^2
\end{eqnarray}

Then, we can compute the $\chi^2$ for the CHARM data:
\begin{equation}
  \chi^2 = \frac{\left(R^e - R^{\rm theo}(\eps_{ee}^{dV},\eps_{ee}^{dA})\right)^2}{({\sigma_R^e})^2}\, ,
\end{equation}
where $R^e$ and $\sigma_R^e$ are defined by the result given in
Eq.~(\ref{Re}).  

\begin{figure}
\includegraphics[width=0.45\textwidth,angle=0]{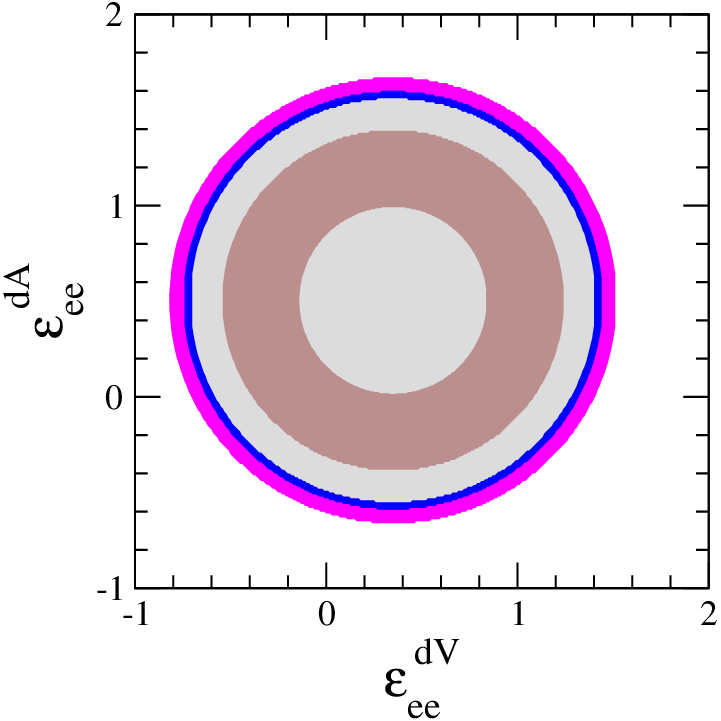}
\caption{ Constraints at 68, 90, 95 and 99\% \CL\ on the neutrino NSI couplings
  $\varepsilon_{ee}^{dV}$ and $\varepsilon_{ee}^{dA}$ from CHARM
  data.}
\label{fig:charm}
\end{figure}

The constraints in the ($\eps_{ee}^{dV}$, $\eps_{ee}^{dA}$) plane at
68, 90, 95 and 99\% \CL\ obtained from this analysis are shown in
Fig.~\ref{fig:charm}. 

\subsection{Combined analysis}
\label{sec:combined}

\begin{figure*}
\begin{center}
\includegraphics[width=0.45\textwidth,angle=0]{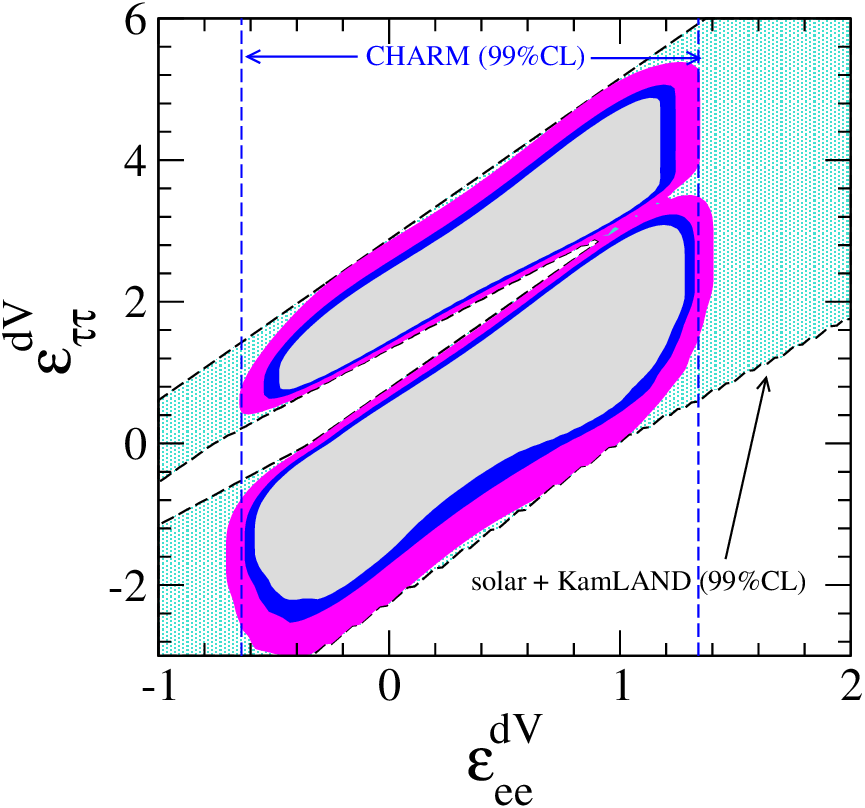}
\includegraphics[width=0.45\textwidth,angle=0]{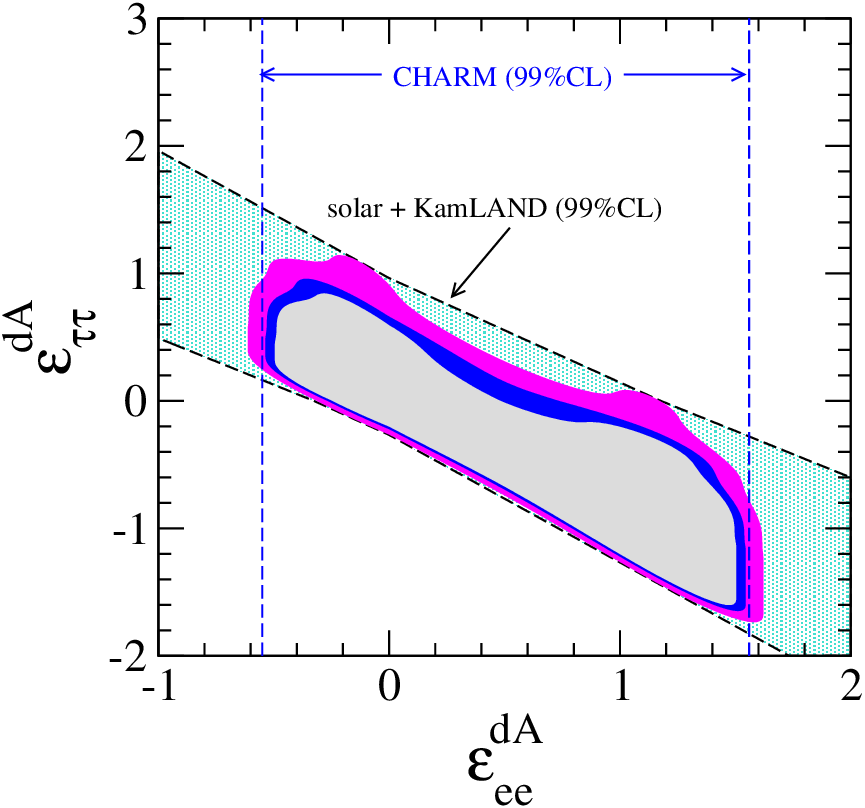}
\caption{ Constraints on the vector (left) and axial-vector (right)
  NSI couplings from our global analysis at 90, 95 and 99\% \CL, and
  from the separate solar plus KamLAND and CHARM data sets (dashed
  lines).}
\label{fig:combined}
\end{center}
\end{figure*}

\begin{table*}
\centering
\caption{New constraints on the vectorial and axial NSI couplings at
  90\% \CL\ obtained from the analysis of CHARM data alone and from our global
  analysis combining CHARM with solar and KamLAND results.}
\vskip .2cm
\label{tab:constraint}
\begin{tabular}{lcc}

  \hline\hline
  \multicolumn{3}{c}{vectorial couplings}  \\
  \hline 
  global &   $-0.5< \varepsilon^{dV}_{ee} < 1.2$  &
  $-1.8< \varepsilon^{dV}_{\tau\tau} < 4.4$  \\[.1cm]
  \hline
  & \multicolumn{2}{c}{one parameter at a time} \\
  \hline 
  CHARM  & $ -0.5 < \varepsilon^{dV}_{ee} < 1.2$ &  \\[.1cm]
  global & $ -0.2 < \varepsilon^{dV}_{ee} < 0.5$ & 
  $ -1.1< \varepsilon^{dV}_{\tau\tau} < 0.4 \, \& \,1.6< \varepsilon^{dV}_{\tau\tau} < 2.2 $ \\[.1cm]
  \hline \\[.1cm]
  \hline \hline
  \multicolumn{3}{c}{axial couplings}  \\
  \hline 
  global &                 $-0.4< \varepsilon^{dA}_{ee} < 1.4$  &
  $-1.5< \varepsilon^{dA}_{\tau\tau} < 0.7$  \\[.1cm]
  \hline
  & \multicolumn{2}{c}{one parameter at a time} \\
  \hline 
  CHARM  & $ -0.4 < \varepsilon^{dA}_{ee} < 1.4$ &  \\[.1cm]
  global & $ -0.2 < \varepsilon^{dA}_{ee} < 0.3$ & 
  $ -0.2 < \varepsilon^{dA}_{\tau\tau} < 0.4$ \\[.1cm]
\hline
\end{tabular}
\end{table*}

In the previous sections we have discussed the sensitivity of solar
experiments, KamLAND and CHARM to the nonstandard interactions
separately.  In this subsection we exploit the complementarity of the
information we can get from the different experiments by combining
CHARM data with our previous results from the analysis of the solar
and KamLAND data. This enables us to obtain stronger constraints on
the NSI couplings.

The regions for the vector (left) and axial-vector (right) NSI
couplings allowed by the global analysis are given in
Fig.~\ref{fig:combined}, where they are also compared with the
constraints coming only from the CHARM data and that from the solar
plus KamLAND data.
First, the bounds obtained from the analysis of CHARM data and shown
in Fig.~\ref{fig:charm}, have been translated into two independent
bounds on $\varepsilon_{ee}^{dV}$ and $\varepsilon_{ee}^{dA}$ (see
vertical bands in Fig.~\ref{fig:combined}).  
The regions allowed by the solar+KamLAND combination (diagonal bands
in the figure) have been derived from the bounds on the non-universal
nonstandard neutrino interactions obtained in
Sec.~\ref{sec:solar-kamland-data}. In particular, the limits on the
vectorial couplings $\eps_{ee}^{dV}$ and $\eps_{\tau\tau}^{dV}$ come
from the allowed values for the effective coupling $\eps'$ in
Eq.~(\ref{eq:eps-lim0}), after using the definition of $\eps'$ of
Eq.~(\ref{eff-coup}) and the 1$\sigma$ allowed region for the
atmospheric mixing angle $\theta_{23}$~\cite{Schwetz:2008er}.
Note the existence of two allowed islands, in correspondence with the
two allowed regions of neutrino oscillation parameters in the presence
of NSI (see upper-right panel at Fig.~\ref{fig:updated}). The lower
one corresponds to the usual LMA solution, while the upper island
comes from the solution in the dark side.
On the other hand, using the average probabilities in
Eq.~(\ref{eq:ncprobs}) we have reanalysed the results obtained for the
effective axial coupling $\eps_A$ in Eq.~(\ref{eq:epsa_bound}). This
results in a constraint for the axial couplings $\eps_{ee}^{d A}$ and
$\eps_{\tau\tau}^{dA}$.
From the two panels of Fig.~\ref{fig:combined} one sees that, as
expected, there is a degeneracy in the determination of the two
vectorial and axial parameters from solar and KamLAND data only.
After the combination with CHARM we break this degeneracy and obtain
the allowed regions shown in color at Fig.~\ref{fig:combined}.

In Table~\ref{tab:constraint} we quote the 90\% \CL ~allowed
intervals for the couplings $\eps_{ee}^{dV}$,
$\eps_{\tau\tau}^{dV}$, $\eps_{ee}^{d A}$ and $\eps_{\tau\tau}^{dA}$
arising from the combined analysis, as taken directly from
Fig.~\ref{fig:combined}.
In order to compare with previous bounds obtained in a
one-parameter-at-a-time analysis~\cite{Davidson:2003ha}, we also
present the results obtained following that approach in our analysis
of CHARM data alone and in the global analysis including solar and
KamLAND data as well.
In this case, we see how the combination with solar and KamLAND data
improves significantly the existing bounds on the electron-neutrino
NSI couplings, obtained using CHARM data only~\cite{Davidson:2003ha}.
Now, concerning the tau-neutrino NSI couplings, we have improved the
existing bounds derived from the invisible decay width measurement of
the Z from LEP data: $|\eps_{\tau\tau}^{dV}| <2.7$, 
$|\eps_{\tau\tau}^{dA}| < 1.8$ \cite{Davidson:2003ha}.

\section{Conclusions}
\label{sec:conclusions}

We have updated the solar neutrino analysis for the case of NSI of
neutrinos with d-type quark by including new solar data from SNO phase
III and Borexino, as well as new KamLAND data and updated solar
fluxes. We have found that the additional dark-side of neutrino
parameter space found in Ref.~\cite{Miranda:2004nb} still survives,
while the previous LMA-0 and LMA-II which were present before have now
disappeared as a result of the new data.
The issue arises of how to lift this degeneracy in future studies.
First we note that, since KamLAND is basically insensitive to matter
effects, it will not help in resolving the degeneracy, as explicitly
verified in Fig.~3 of Ref.~\cite{Miranda:2004nb}.
Next comes improved solar neutrino data. The form of the
expected neutrino survival probability shown in Fig.~\ref{fig:pee}
suggests that the best region to discriminate the degenerate
solution from the normal one is the intermediate energy solar
neutrino region, where the relevant experiments are
Borexino~\cite{Arpesella:2008mt,Galbiati:2008zz,Collaboration:2008mr}
and KamLAND-solar~\cite{Nakamura:2004cb}, as well as the low energy
threshold analysis expected from Super-K-III and
SNO~\cite{Takeuchi:2008zz,SNO-LETA}.  

Further information relevant to lift the degeneracy may come from
atmospheric and laboratory data. Indeed the LMA-D solution induced
by non-standard interactions of neutrinos with quarks may become
inconsistent with atmospheric and laboratory
data~\cite{Miranda:2004nb, Friedland:2004ah}.
As noted in Ref.~\cite{Miranda:2004nb} currently it is not.
Should the situation change with improved data we note that still the
degeneracy will not disappear 
since the NSI couplings may affect not only
down-type quark species but also up-type quarks and/or electrons. The
analysis in this case would introduce new parameters.
Therefore, we conclude that the neutrino oscillation interpretation of
solar neutrino data is still fragile with respect to the presence of
nonstandard interactions.

In summary,  we have studied the limits on the
non-standard interaction couplings $\eps_{ee}^{d V}$,
$\eps_{e\tau}^{dV}$, $\eps_{\tau\tau}^{dV}$, $\eps_{ee}^{d A}$ and
$\eps_{\tau\tau}^{dA}$ from present neutrino data.  Thanks to the combination of solar and KamLAD neutrino data with the results from the CHARM experiment, we have given improved
bounds on the vector and axial NSI couplings involving electron and tau neutrino
interactions on down-type quarks.

\section*{Acknowledgments}

Work supported by Spanish grants FPA2008-00319/FPA and
PROMETEO/2009/091.  OGM was supported by CONACyT-Mexico and
SNI. M.A.T. is supported by the DFG (Germany) under grant SFB-676. FJE
thanks Cinvestav for hospitality when part of this work was performed.

\renewcommand{\baselinestretch}{1}

%\bibliographystyle{h-physrev4} 
% \bibliography{valle-ref,bibt3,morisi-ref}%%,soko,snova,nu-rev06,parke-ref}
%\bibliography{valle-ref}%%,soko,snova,nu-rev06,parke-ref}

\end{document}